%% file: main.tex
\documentclass[sigconf]{acmart}

\AtBeginDocument{%
  \providecommand\BibTeX{{%
    \normalfont B\kern-0.5em{\scshape i\kern-0.25em b}\kern-0.8em\TeX}}}

\setcopyright{acmcopyright}
\copyrightyear{2023}
\acmYear{2023}
\setcopyright{acmlicensed}\acmConference[ETRA '23]{2023 Symposium on Eye Tracking Research and Applications}{May 30-June 2, 2023}{Tubingen, Germany}
\acmBooktitle{2023 Symposium on Eye Tracking Research and Applications (ETRA '23), May 30-June 2, 2023, Tubingen, Germany}
\acmPrice{15.00}
\acmDOI{10.1145/3588015.3589841}
\acmISBN{979-8-4007-0150-4/23/05}

\usepackage{subfigure}
\usepackage{enumitem}
\usepackage{hyperref}
\usepackage{tabularx}

\begin{document}

\title{Reading Strategies for Graph Visualizations that Wrap Around in Torus Topology}
% \author{
% Kun-Ting Chen\textsuperscript{1}, 
% Quynh Quang Ngo\textsuperscript{1},
% Kuno Kurzhals\textsuperscript{1},
% Kim Marriott\textsuperscript{2},
% Tim Dwyer\textsuperscript{2},
% Michael Sedlmair\textsuperscript{1},
% Daniel Weiskopf\textsuperscript{1}\\
% {\large\normalfont 
% \textsuperscript{1} University of Stuttgart, Germany,
% \textsuperscript{2} Monash University, Australia
% }
% \\
% {\large\normalfont 
% \{kun-ting.chen, Quynh.Ngo, michael.sedlmair, Daniel.Weiskopf\}@visus.uni-stuttgart.de, 
% Kuno.Kurzhals@vis.uni-stuttgart.de,
% \{kim.marriott,tim.dwyer\}@monash.edu
% }
% }

\author{Kun-Ting Chen}
\orcid{0000-0002-3217-5724}
\affiliation{%
  \institution{University of Adelaide}
  % \city{Stuttgart}
  % \state{Baden Württemberg}
  \country{Australia}
  % \postcode{70569}
}
\affiliation{%
  \institution{University of Stuttgart}
  % \city{Stuttgart}
  % \state{Baden Württemberg}
  \country{Germany}
  % \postcode{70569}
}
\email{Kun-Ting.Chen@adelaide.edu.au}

\author{Quynh Quang Ngo}
\orcid{0000-0001-5254-1480}
\affiliation{%
  \institution{University of Stuttgart}
  % \city{Stuttgart}
  % \state{Baden Württemberg}
  \country{Germany}
  % \postcode{70569}
}
\email{Quynh.Ngo@visus.uni-stuttgart.de}

\author{Kuno Kurzhals}
\orcid{0000-0003-4919-4582}
\affiliation{%
  \institution{University of Stuttgart}
  % \city{Blacksburg}
  % \state{VA}
  \country{Germany}
  % \postcode{24060}
}
\email{Kuno.Kurzhals@vis.uni-stuttgart.de}

\author{Kim Marriott}
\orcid{0000-0002-9813-0377}
\affiliation{%
  \institution{Monash University}
  % \city{Melbourne}
  % \state{VIC}
  \country{Australia}
  % \postcode{3145}
}
\email{Kim.Marriott@monash.edu}

\author{Tim Dwyer}
\orcid{0000-0002-9076-9571}
\affiliation{%
  \institution{Monash University}
  % \city{Melbourne}
  % \state{VIC}
  \country{Australia}
  % \postcode{3145}
}
\email{Tim.Dwyer@monash.edu}

\author{Michael Sedlmair}
\orcid{0000-0001-7048-9292}
\affiliation{%
  \institution{University of Stuttgart}
  % \city{Stuttgart}
  % \state{Baden Württemberg}
  \country{Germany}
  % \postcode{70569}
}
\email{Michael.Sedlmair@visus.uni-stuttgart.de}

\author{Daniel Weiskopf}
\orcid{0000-0003-1174-1026}
\affiliation{%
  \institution{University of Stuttgart}
  % \city{Stuttgart}
  % \state{Baden Württemberg}
  \country{Germany}
  % \postcode{70569}
}
\email{Daniel.Weiskopf@visus.uni-stuttgart.de}

\renewcommand{\shortauthors}{Chen and Ngo, et al.}

\newcommand{\rev}[1]{\textcolor{black}{#1}}
\newcommand{\kt}[1]{\textcolor[RGB]{255, 165, 0}{kt: #1}}
\newcommand{\qn}[1]{\textcolor[RGB]{255, 0, 0}{QN: #1}}

\newcommand{\tnocontext}{\textsc{NoContext}}
\newcommand{\tfullcontext}{\textsc{FullContext}}
\newcommand{\tpartialcontext}{\textsc{PartialContext}}
\newcommand{\tnodelink}{\textsc{NoTorus}}
\newcommand{\deasy}{\textsc{Small}}
\newcommand{\dmedium}{\textsc{Medium}}
\newcommand{\dhard}{\textsc{Large}}

\newcommand{\tshortestpath}{\textsc{ShortestPath}}
\newcommand{\tneighbours}{\textsc{Neighbors}}
\newcommand{\tnodecount}{\textsc{NodeCount}}
\newcommand{\tlinkcount}{\textsc{LinkCount}}

\begin{abstract}
We investigate reading strategies for node-link diagrams that wrap around the boundaries in a flattened torus topology by examining eye tracking data recorded in a previous controlled study.
Prior work showed that torus drawing affords greater flexibility in clutter reduction than traditional node-link representations, but impedes link-and-path exploration tasks, while repeating tiles around boundaries aids comprehension.
However, it remains unclear what strategies users apply in different wrapping settings. This is important for design implications for future work on more effective wrapped visualizations for network applications, and cyclic data that could benefit from wrapping.
We perform visual-exploratory data analysis of gaze data, and conduct statistical tests derived from the patterns identified.
Results show distinguishable gaze behaviors, with more visual glances and transitions between areas of interest in the non-replicated layout. Full-context has more successful visual searches than partial-context, but the gaze allocation indicates that the layout could be more space-efficient.

\end{abstract}






\begin{teaserfigure}
    \centering   
     \includegraphics[width=1.0\textwidth]{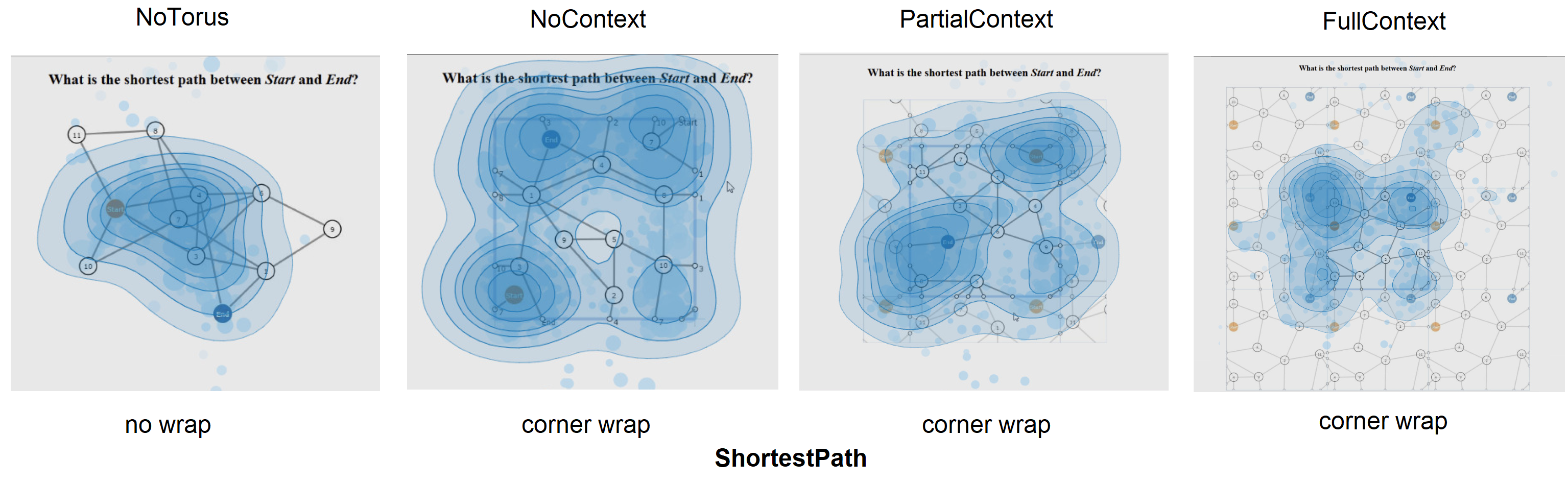}
    \caption[]{Density maps of time-aggregated fixation distributions overlaid in the original image of four network layouts of a given graph. Higher saturation indicates denser fixation distribution over regions of the image. Participants were asked to identify the unique shortest path between the start (orange) and the end (dark blue) node. The radius of the light blue dots indicate the fixation duration. 
    % Each stimulus shows the same example of a medium-sized graph where the unique shortest path is wrapped around in top-bottom, left-right, or both directions (corner wrapping), or without any wrapping (\tnodelink). Corner wrapping is compared for different layouts with varying degree of repeated visual context: \tnocontext, \tpartialcontext, and \tfullcontext.
    \label{fig:teaser}
    }
    \Description[]{}
\end{teaserfigure}

\maketitle
% \includegraphics[width=1\textwidth]{figures/doughnets_case_study.png}
% \caption{figure caption}
% \Description{figure description}
% \label{fig:teaser}

\input{sections/introduction}
\input{sections/background}
\input{sections/study}
\input{sections/analysis}
% \input{sections/discussion}
\input{sections/conclusion}

\begin{acks}
The authors like to thank Hong Gui for the paper writing suggestions, as well as all the anonymous reviewers for their helpful comments on this paper.
This research is funded by the Deutsche Forschungsgemeinschaft (DFG, German Research Foundation) – Project-ID 251654672 – TRR 161 and under Germany’s Excellence
Strategy – EXC 2120/1 – 390831618.
\end{acks}

% \begin{acks}
% \end{acks}

\bibliographystyle{ACM-Reference-Format}

\bibliography{references}

% \appendix
% \input{sections/appendix.tex}
\end{document}

%% file: sections/introduction.tex
\section{Introduction}
% \begin{figure}
%     \centering
    
%      \includegraphics[width=0.7\textwidth]{figures/torus.pdf}
    
%     \Description[]{}
%     \caption[]{Illustrative drawing of the wrapping in a 2D flattened torus space considered in this paper: image from left to right shows a 3D torus shape; a resulting flattened torus where data points can be mapped (projected) to a bounded Euclidean space, 
%     The space is glued with connectivity at the boundaries of such projection connecting vertically between top and bottom, and horizontally between left and right; the third and fourth column shows tile replication which provides a continuous view of graphical elements wrapped across the boundary.
%     % and have different connectivity at the boundaries of such projection; the third and fourth column shows tile replication which provides a continuous view of graphical elements wrapped across the boundary.
%     \label{fig:torus}
%     }
% \end{figure}

\begin{figure}
    \centering
    \subfigure[\tnodelink]{
    \includegraphics[width=0.21\textwidth]{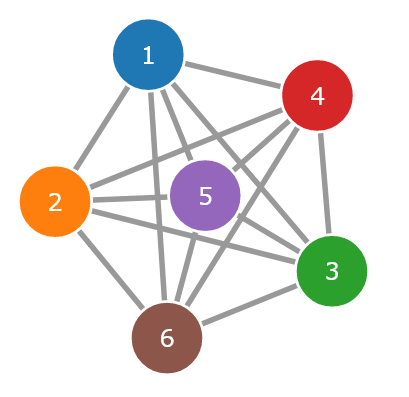}
    }
    \subfigure[\tnocontext]{
    \includegraphics[width=0.21\textwidth]{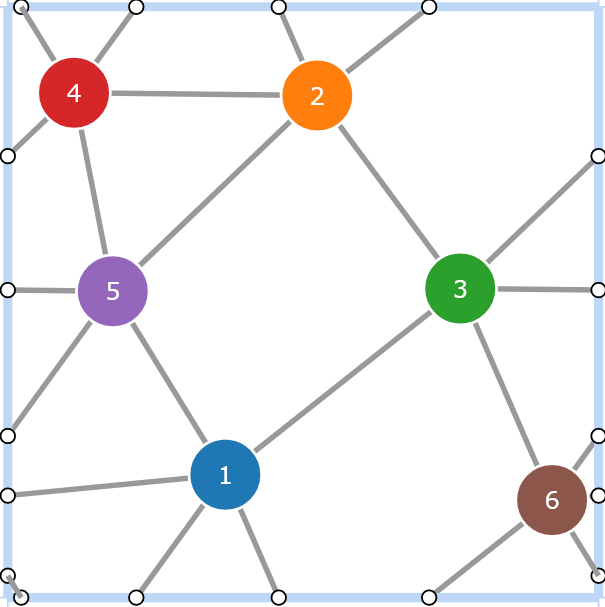}
    }
    \subfigure[\tpartialcontext]{
    \includegraphics[width=0.21\textwidth]{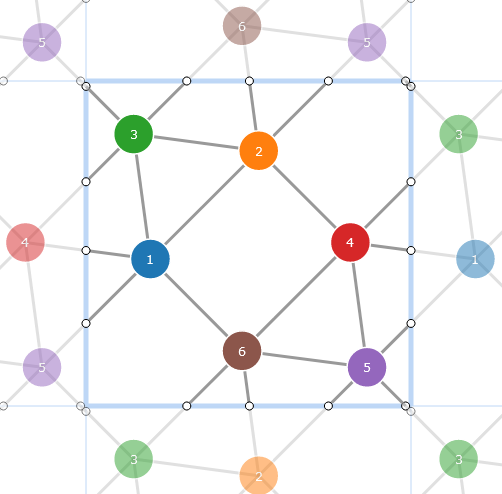}
    }
    \subfigure[\tfullcontext]{
    \includegraphics[width=0.21\textwidth]{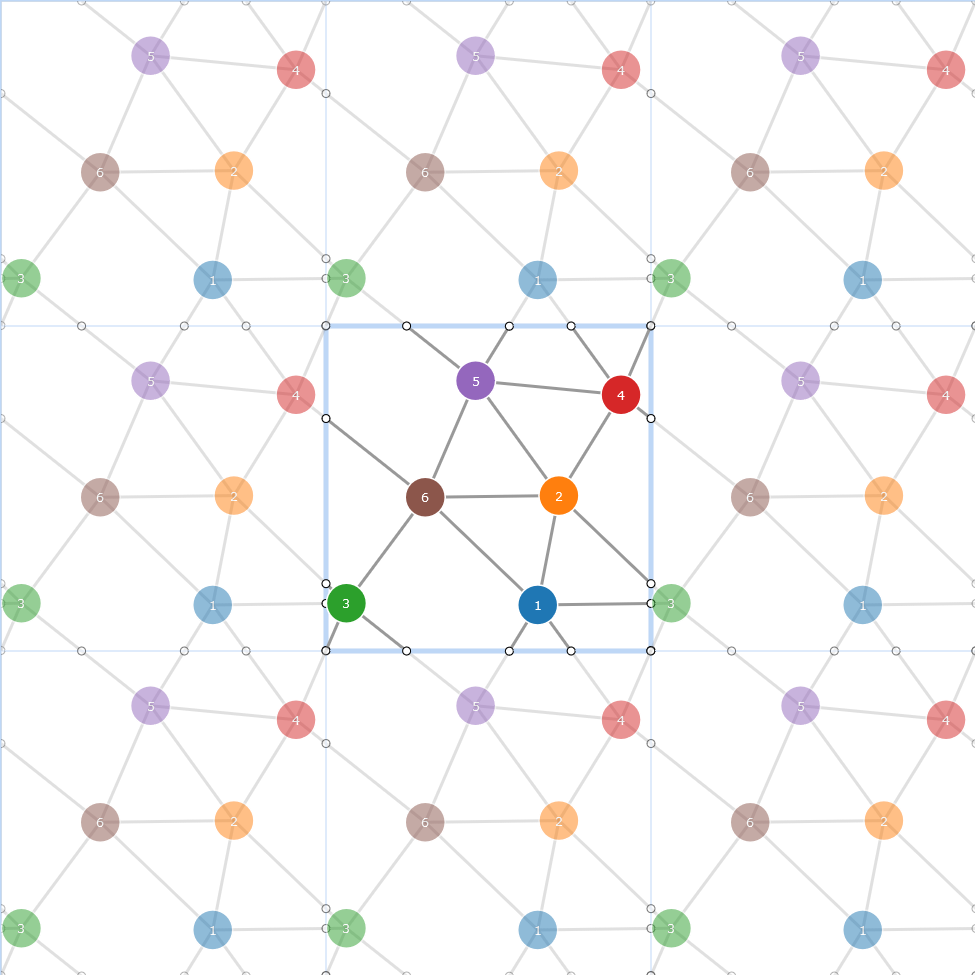}
    }
    \caption{
    The complete k6 graph drawn using a traditional node-link layout (\tnodelink) and three different techniques in a 2D flattened torus space considered in this paper. 
    % data points are mapped (projected) to a bounded Euclidean space. In the torus configuration, the boundary is glued with links wrapped vertically between top and bottom, and horizontally between left and right (\tnocontext). The third and fourth columns show partial (\tpartialcontext) or full tile replication (\tfullcontext), providing a continuous view of graphical elements wrapped across the boundary.
    }
    \vspace{-1em}
    \label{fig:k6_four_conditions}
\end{figure}

Traditional 2D network visualizations are designed to be viewed on a printed page or standard screen. Recently, ~\cite{chen2020doughnets} have shown that by laying out a network on a flattened 2D torus topology, the layout algorithm has more flexibility in relaxing node positions, thereby improving graph layout aesthetics. 
However, a downside of such representations is the visual cut of network links introduced at the boundaries of the display that requires mental wrapping to bridge the cuts~\cite{chen2021rotate} (\autoref{fig:k6_four_conditions}(b)).
To address this problem, repeated tiles of the original diagram can be drawn around the boundaries of the display, providing partial (\tpartialcontext, see \autoref{fig:k6_four_conditions}(c)) or even full context (\tfullcontext, see \autoref{fig:k6_four_conditions}(d)) in the layout. 
To further mitigate the impact of such visual cuts, interactive panning~\cite{chen2021sa,chen2021rotate} or a combination of panning and rotation~\cite{chen2022gan} can be used to center the content of interest. 
% improves task performance of cyclic time series data~\cite{chen2021rotate} and inspection of high-level network structures in node-link diagrams~\cite{chen2022gan}. 
Yet, interactivity in visualization requires human intervention, suffers from increased mental effort~\cite{munzner2014visualization}, and is unavailable in public displays or printouts. 

In this work, we focus on understanding users' reading strategies for stationary wrapped layouts. The controlled user study by 
\citet{chen2020doughnets} showed that \tfullcontext\ was the best torus layout in terms of error rate, speed, and subjective preference, and as good as traditional node-link without link wrapping (\tnodelink).
% \tfullcontext\ was found to performs as good as the technique without wrapping.
% Here we explore a different solution that still uses node-link diagrams
% %
% %
% ~\cite{chen2020doughnets} recently evaluated in an eye tracking study a technique  using a force-directed approach which maps arbitrary network structures embedded on a toroidal surface to diagrams with wraparound links across the boundaries of the display. %\qn{long sentences, and if it was an eye tracking study, what the point of having this paper?}
% %
% Compared to a traditional layout, a toroidal topology diagram allows for links to be wrapped horizontally and vertically around the boundaries of the display, so the layout algorithm has more flexibility in relaxing the node positions to better capture the data's underlying relationships~\cite{chen2022s}.
\rev{In \citet{chen2020doughnets}'s study, although participants' visual focus was collected using an eye tracker, the analysis of their study primarily centered on task effectiveness not the visual search behavior, nor did they publish any gaze analysis results.
Therefore, this leaves open questions related to aspects such as the mental process and different reading strategies required to perform task over a wrap-around representations. In this paper, we publish the eye tracking dataset\footnote{https://osf.io/au3bj/} and report its analysis addressing these questions.} 

To fill this gap, we perform an exploratory analysis of this eye tracking data and conduct statistical tests derived from the gaze patterns identified.
Our results show that there are distinguishable reading strategies for each layout condition, varying by tasks. In general, when content of interest (e.g., unique shortest path) wraps around the boundary, the \tnocontext{} layout leads to a strategy with more frequent long-range glancing after immediately fixated areas, and has significantly more transitions between two or more clusters of fixation points than other layout conditions.
As implication for future design, we thus suggest to automatically center content of interest to avoid  visual cuts.
In \tpartialcontext, frequent visual searches over redundant areas are identified. 
In \tfullcontext, the visual search concentrates on one or two complete shortest paths, leading to reduced task-completion times; however, the large screen space is quite inefficiently used according to our analysis of scanpaths. 
% The results in analysing the eye tracking data show that there are distinguishable graph reading strategies in each layout condition, varying across tasks. 
% By examining the user reading strategies in different wrapping conditions, we found density of visual attention for no-wrapping in a torus-based layout is similar to none-torus layout (having one cluster of fixations). 
% When content of interest wraps around the boundaries, all torus drawings result in two or more clusters, while with repeated tiles it significantly reduces inter-cluster visual search, and thus reduce cognitive effort.
% This finding suggests future design of torus-based layout should avoid the visual cut, while leveraging the advantages of improved aesthetics for improving graph readability. 
%
%\qn{TODO: revise below, not all results details should be lsited here}
%We find that there is a higher probability for horizontal and vertical visual scan than corner scanning directions. 
%

Our findings also indicate that the three-by-three tiled display in \tfullcontext{} is still necessary, %(rejecting hypothesis of RQ1), 
but not a full level of replication at the corners has to be utilized, suggesting a more compact representation of a full-context layout. 
Our results complement existing task performance analyses and offer implications for future torus-based 
%layouts a for general wrapped 
data visualizations.
% where stationary layouts with repeated tiled display is utilized~\cite{chen2022s}.
% The results comparing different layouts show that the average fixation duration and frequent eye movements (i.e., AOI transitions) are significantly smaller in full-context and traditional node-link, compared with none-replicated layout and partially replicated layout, suggesting more mental effort is involved in none-replicated and partially replicated layouts.

%% file: sections/background.tex
\section{Related Work}
\label{sec:background}
% \cite{chen2022s} has recently investigated the utility of network visualisations based on projecting the surface of a cylinder, sphere, or torus onto a flat 2D plane.
% The authors presents a design space generalising design dimensions of topology, wrapping and repeated tiled displays in unifying a class of wrappable data visualisations. 
% Crucially, for wrapped visualisations the viewer can utilize additional context while observing visual cuts (e.g., links between two nodes split across the boundaries) that ``wraps'' around the plane. 
% Providing replicated context have previously been found to be of benefit compared with the case without any context in link and path following tasks~\cite{chen2020doughnets}. 
% Here we investigate the reading strategies of various levels of replicated context.

%
% Towards this end, we intentionally find from our eye-tracking data the potential gaze analytics metrics and graph reading strategies that match task performances in reading different node-link designs. 
%
% In this section, we survey recent related work deriving such metrics and strategies from eye-tracking data. 

% \subsection{Visualisations that wrap around in topologies}

% \textbf{Graph reading strategies extracted from eye-tracking data}
Graph drawings have been previously analyzed with eye tracking studies and revealed different reading strategies depending on network layout~\cite{pohl2009comparing,burch2013visual}, tasks~\cite{huang2009graph}, quality criteria~\cite{Huang2013}, and coordinated views~\cite{chang2017evaluating}. 
% These findings are low level and provide guidance to analyzing torus-based layout. 
We extend this line of work by analyzing gaze data characteristics for a new network layout, that is, wrapped layouts of node-link diagrams. 

Eye tracking data has been reported to be useful in identifying graph reading strategies~\cite{Tanja2017NodeLink}. 
%
%For graph layout, 
\citet{GraphLayoutEffectEyeTracking2007, Huang2013} 
observed eye movement to extract graph reading behavior given different crossing angles, and deductively recommended designing crossing angles approximately $90$ degrees to achieve optimal task performance. %performed an eye tracking study in understanding how crossing angles in the graph layout influence eye movements and identifying geometric-path searching tendencies from the user.
%
%~\cite{Huang2013} looked at eye tracking video data and identified that the edge crossing angle and the distracting branches of edges confused eye movement, not the crossings. 
%
However, he mainly based his eye tracking data analysis on observing eye movement videos, whereas visual analytics could help gain deeper insights~\cite{Andrienko:2012:VAM,kurzhals2017task}. %\textcolor{red}{QN: Huang has two papers here, 2007, and 2013. the 2013 one use videos, not sure about 2007. TODO}
\citet{pohl2009comparing} compared graph readability strategies in force-directed, orthogonal, and hierarchical graph layouts with eye tracking. 
They found that, for path-following tasks, orthogonal and force-directed layouts come with less link tracing effort than hierarchical layouts due to more visual attention found in line crossings in hierarchical layouts.  
\citet{EyeTrackingTree2011} utilized an analysis based on areas of interest (AOIs) to identify the differences in the readability of three tree visualization layouts. 
The time to cross-check between AOIs explained why the radial layout was less efficient in performing the given task. 
\citet{burch2013visual} further analyzed the eye tracking part of this study in more detail in order to identify reading strategies by employing exploratory visual analytics; we follow a similar approach in our paper, but for a different kind of layouts. %user study.
%
%
%~\cite{MichaelBurch2014DynamicGraph} (not relevant)
%
\citet{DisplayGroupInfoNodeLink2014} performed a small-scale eye tracking study to evaluate four different augmentation visualization strategies on top of a node-link diagram for group-visualization-related tasks. 
%
%
%\cite{netzel2017user} visually compared scanpaths of multiple users reading strategies for path following in metro maps and evaluate the respective user performance.\textcolor{red}{QN: what is the eye tracking revealing strategies reading? What are implication?} 
%
%~\cite{blascheck2018NodeLinkGraphLayout} \textcolor{red}{QN: todo - read this and write} . \qn{sorry, I don't have time to read this paper. :)}
%
Variations in task solution strategies for path following tasks are presented in metro maps reading strategies~\cite{netzel2017user,kumar2019clustered}.
%
%Kurzhals et al. proposed to use space time cube (https://vbruder.github.io/talk/2019-etra/), and Burch et al.~\cite{MichaelBurch2014DynamicGraph} propose to use dynamic network visualization for scanpaths analysis tasks.
%
%\textcolor{red}{QN: write few more sentences to relate back to our work.}

While all of the aforementioned work focused on visual guidance aspects for traditional node-link diagrams with no wrapping, we focus on studying wrapped layouts.
%differ and finding potential suggestions for improvement of the design. 

% \subsection{Gaze analytics metrics in graph reading}
% Several studies used gaze metrics to interpret graph reading strategies performing a specific task. 
% %
% One common interpretation for path tracing between source and destination over a metro map:
% High fixation duration means typically that a participant investigated an area more carefully, since the task has demanded it or due to a high degree of complexity. 
% %
% Lower values, occurring in a restricted area in combination with a small average saccade length, could be an indicator for stress~\cite{netzel2017user,netzel2014comparative}. 
% %
% This interpretation is also used to explain their task performance.  
% %
% Saccade transition frequencies are also used to indicate higher visual search effort for graphs [x]. \textcolor{red}{QN: why [x]?}

%% file: sections/study.tex
\section{Eye Tracking Study}
In the following, we briefly review the main aspects of the previous experiment conducted by~\citet{chen2020doughnets}. Our work focuses on analyzing eye tracking data that was collected in this study, while the earlier publication focused on task performance. The following description thus focuses primarily on the eye-tracking part, as well as general aspects relevant to our analysis. For more details, we point the reader to~\citet{chen2020doughnets}.  

%as far as they are relevant for this paper.

\subsection{Design and Procedure}
The study used a repeated-measure design, in which each participant performed a set of tasks on the four different graph layout conditions shown in~\autoref{fig:teaser}.
% The study was based on a repeated-measure design in within subjects style. 
Eye movement data of 19 participants was collected. Gaze sequence data was dropped if it was below a threshold of 70\% gaze samples detected with at least one eye.
% This ensures gaze data quality
% (with at least one eye detected according to the Tobii Pro eye tracker. 
This resulted in valid data from 13 participants.
A Tobii Pro X3-120 eye tracking system was used to record eye movement data at a sampling rate of 120\,Hz. Participants sat in front of a Dell 22-inch LCD monitor at a distance given by the calibration feature of the eye tracking system. 
%
%\qn{users's demographic? or link to the previous paper?}
%

Before conducting the experiment, subjects were asked to fill out a questionnaire about personal details and an online tutorial was provided to explain the torus-based network diagrams used in the study. 
There were four training trials for each layout condition right before the recorded trial.
Each trial began with question text located at the top. Once a participant hit the start button, an image stimulus was loaded below the question and a timer started. The image was shown for a maximum duration of 30 seconds. A progress bar appeared below the stimulus 10 seconds before the timeout. 
Once the button was hit, the image disappeared and the question showed up again, with multiple-choice options for answering the question. 
% Finally, multiple-choice radio buttons or a slider appeared right below the question text to obtain the participant's answer.

\subsection{Datasets and Tasks}
In the actual study, stimuli were given in the form of  \tnodelink{} \rev{(i.e., a standard force-directed node-link representation of a network)}, \tnocontext{} \rev{(i.e., a torus drawing without any contextual tiling around the boundary)}, \tpartialcontext{} \rev{(i.e., a torus drawing with partial contextual tiling around the boundary to aid comprehension of the link wrapping)}, and \tfullcontext{} layouts \rev{(i.e., a torus drawing with full three-by-three contextual tiling around the boundary  to aid comprehension of the link wrapping)}. An example graph is shown in~\autoref{fig:scanpath_strategies}. 
The fully replicated network layouts have 8 repeated cells (top, right, bottom, left, and four corners). Partially replicated layouts show one third of the fully replicated layouts around the boundaries of the center cell.  
% For partially and fully replicated layouts, the center cell has a slightly higher opacity than the adjacent and corner cells.

The tasks were (1) to locate the shortest path of a set of orange and blue nodes (\tshortestpath{}) with text labels indicating start and end node; (2) to identify all neighbors of an orange node (\tneighbours{}); and (3) to count the total number of nodes (\tnodecount{}) and links (\tlinkcount{}). 
The latter task was chosen because the link wrapping had to be understood and a strategy had to be applied to answer correctly. The applied strategies are of special interest in this paper.

Note that the diagrams are scaled to the same size in the figures in this paper for a clear illustration of visual scanpaths and density maps, but in the eye tracking experiments, stimuli were shown to participants so that the central cell was the same size as the non-replicated drawing; i.e., the area of fully replicated layout drawings (975\,$\times$\,975) shown to study participants were nine times that of non-replicated drawings (325\,$\times$\,325).

\subsection{Previous Results}
% In the previous publication, ~\citet{chen2020doughnets} demonstrated that for networks with up to 15 nodes and 36 links, the torus drawing %leads to improved 
% improves layout quality metrics, including fewer link crossings, smaller stress, more equal link length, and less deviation in incidence angle between links where they connect to a node.
%, compared with none-torus based representations. 
%
% Detailed statistical tests showed significant differences between
% layouts, i.e., the task could be answered much faster in \tfullcontext{} layout and traditional node-link layouts than none replicating layout and partially replicated layouts for shortest path and counting link tasks.
%
Overall the task performance results suggested that fully replicated and traditional layouts outperform \tnocontext{} and \tpartialcontext{} for link-and-path-following tasks in accuracy, completion time, and the subjective user rank.
Their qualitative user feedback reported that while major of participants favored \tfullcontext{} for taking an entire overview of the network, others mentioned the extra replication looked redundant and preferred the clean representation of the \tnocontext{} layout. 
Towards a better understanding of user reading strategy of each layout, we perform an eye tracking data analysis.
% However, the authors did not explain further the difference between none and replicated layouts in terms of qualitative user strategies. Instead, they conducted a second study to augment the \tnocontext{} layout with interactive wrapping. 

%% file: sections/analysis.tex
\section{Analysis of reading strategies}
\begin{figure*}
    \centering    
     \includegraphics[width=1.0\textwidth]{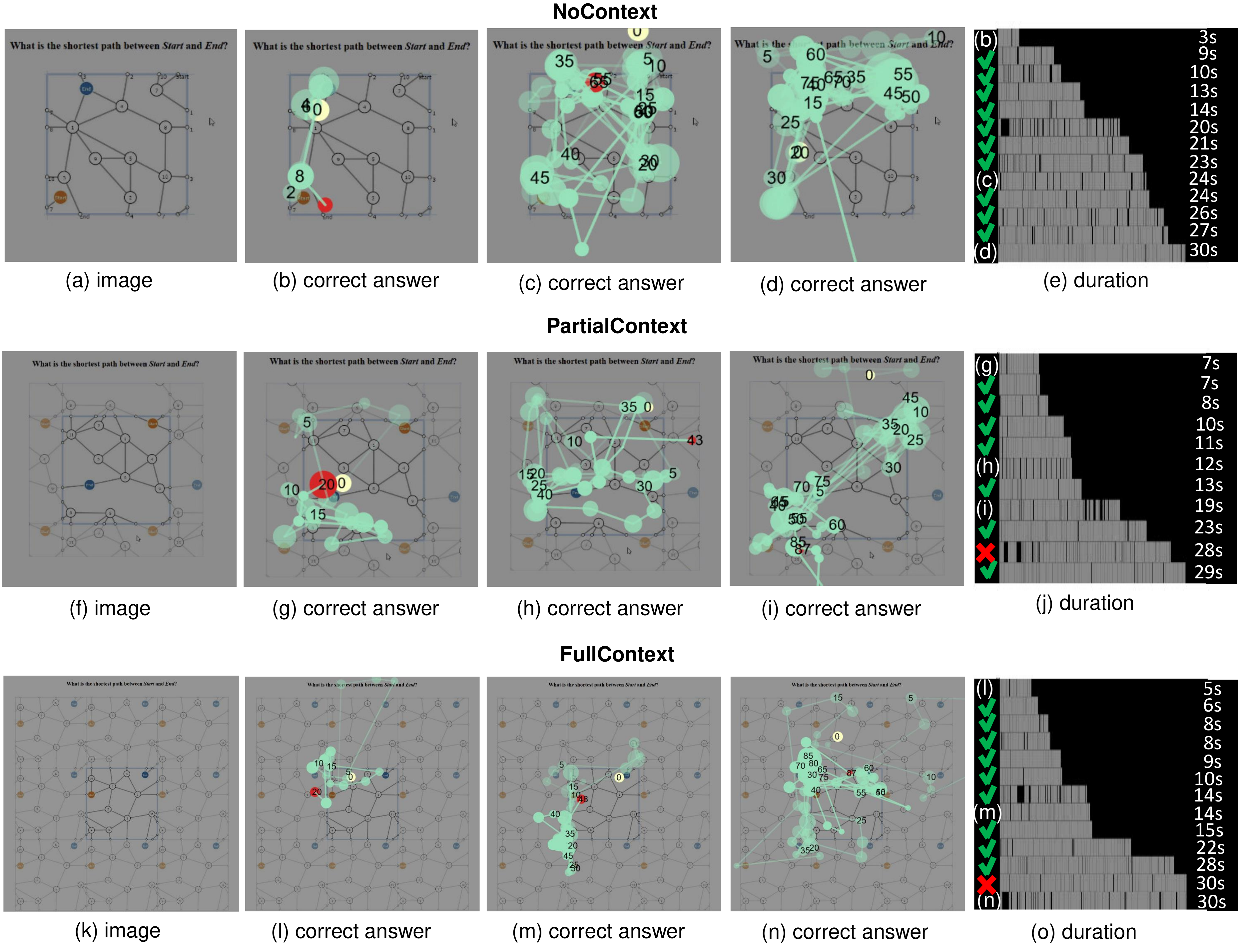}
    \Description[]{}
    \caption[]{Visual scanpath overlaid on images of three torus-based layouts of a representative stimuli for \tshortestpath{}. 
    The number and radius indicate the fixation sequence and its duration, respectively. 
    Increasing saturation of a link encodes temporal progression of a scanpath.
    The yellow and red dots indicate the beginning and the end of a scanpath.   
    Scarf plots ordered participants by increasing task completion time, annotated with task correctness in a green tick (for a correct answer) or a red cross (for an incorrect answer).
    % Images are ordered by increasing task completion time from the left to the right.    
    % \qn{color is inaccessible?, and the background in timelines better to be gray like in stimuli's} 
    \label{fig:scanpath_strategies}
    }
\end{figure*}

\begin{figure}
    \centering    
     \includegraphics[width=1.0\columnwidth]{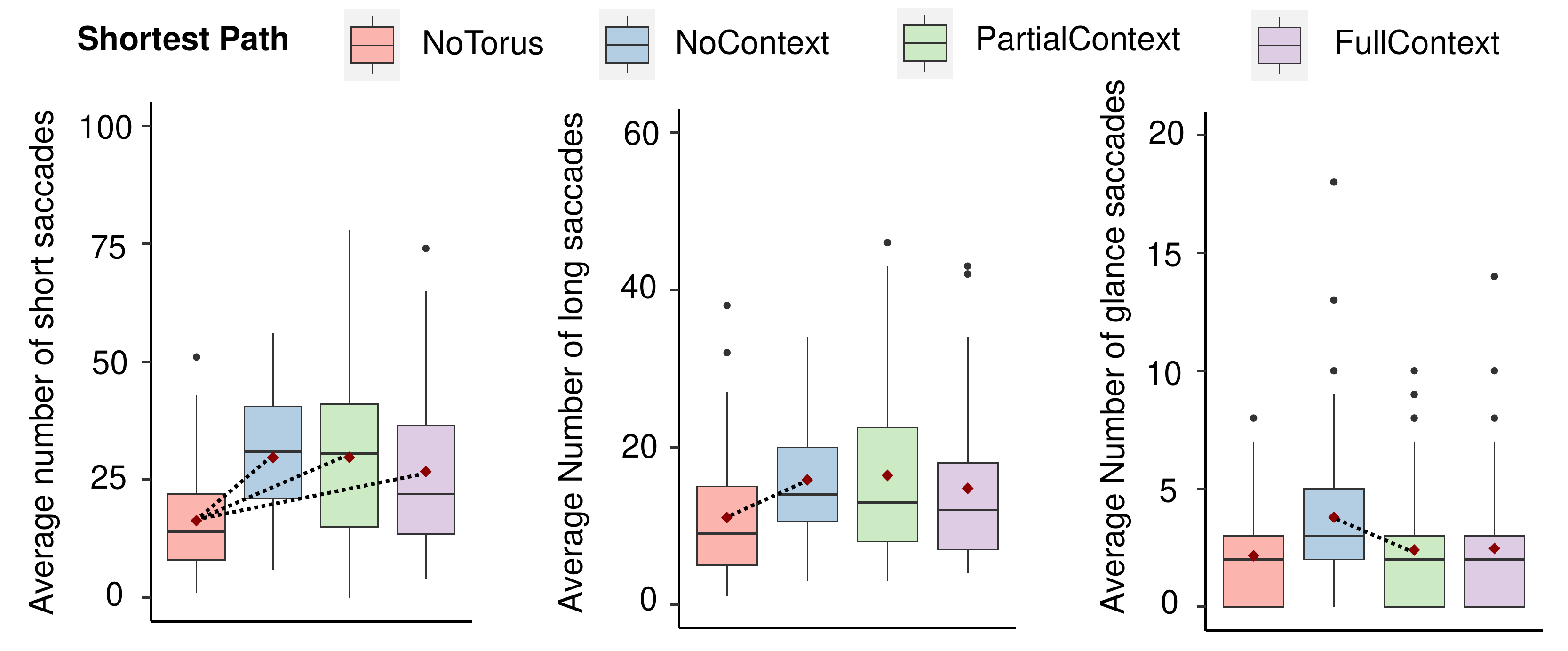}
    \Description[]{Box plot of number of short saccades, long saccades, and glance saccades across participants for each layout condition. The short saccade cutoff length is 100 pixels, which is less than one third of the width of the central tile.
    }
    \caption[]{Box plot of average number of short saccades, long saccades, and glance saccades per participant across medium and large graphs for each layout condition shown in~\autoref{fig:scanpath_strategies}. The short saccade cutoff length is 100 pixels, which is about one third of width of the central tile. Dotted lines connecting average indicates significant difference. 
    }    
    \label{fig:number-of-saccades}
    \vspace{-1em}
\end{figure}
% \textcolor{red}{QN: what are the hypotheses?}
% \textcolor{red}{QN: what are the tasks specifically for the users?}
% \textcolor{red}{QN: what are interesting results from reading/analyzing the eye-tracking data. Give some hints here}
Our overarching goal is to understand visual scanning strategies that participants apply to perform graph exploration tasks with layouts that wrap around in different levels of repeated tiled display. 
As the focus of the tasks was on link-and-path following, we define a reading strategy as a sequence of frequently occurring spatial fixation points in an image stimulus out of correctly performed samples. 
A similar data model has also been used by \citet{burch2013visual}.
% We adopt standard eye tracking data analysis approaches of graph layouts reported in the literature~\cite{kurzhals2017task}. hey, could you check the last sentence in 4.3? NoCONTEXT has more ... in NOCONTEXT?  One of them is NOTORUS instead? I'm not sureShould we delete the in NOCONTEXT? Overall, it feels like we had only results comparing NOCONTEXT with other layouts.
The fixation identification parameters are based on a dispersion-based algorithm (I-DT)~\cite{salvucci2000identifying}. 
We choose dispersion as 130 pixels by the sum of a bounding box $([\max(x) - \min(x)] + [\max(y) - \min(y)])$ for a cluster of gaze points, and minimum time window as 80 milliseconds. This results in about 1--2 fixations over a node in the network diagram we presented to participants.
% For example, in stimuli shown in~\autoref{fig:teaser}, the distribution of the number of fixations is (M, STD).

We first perform exploratory data analysis to understand graph reading strategies. 
We visually inspect overall fixation distributions of each stimuli, across all participants, to gain an overview of the gaze data. 
We then inspect the scanpaths of each individual participant, analyzing scanpath similarity, and summarize visual investigation patterns found for each layout and task. 
Secondly, informed by the patterns identified, we perform comparative analysis with statistical inference to investigate the difference in gaze metrics regarding layout conditions. 
We use a visual gaze tracking analysis toolkit, called Gazealytics~\cite{chen2023gazealytics}~\footnote{https://github.com/gazealytics/gazealytics-master}, to perform both of such exploratory and comparative analysis. 

The analysis and report is based on a representative stimuli. Similar method has been used to report an existing eye tracking study by~\citet{burch2013visual}.

In the following, we present results of the \tshortestpath{} task. The results of \tlinkcount{} tasks are available in the supplementary material.

\begin{figure*}
    \centering
    \includegraphics[width=0.9\textwidth]{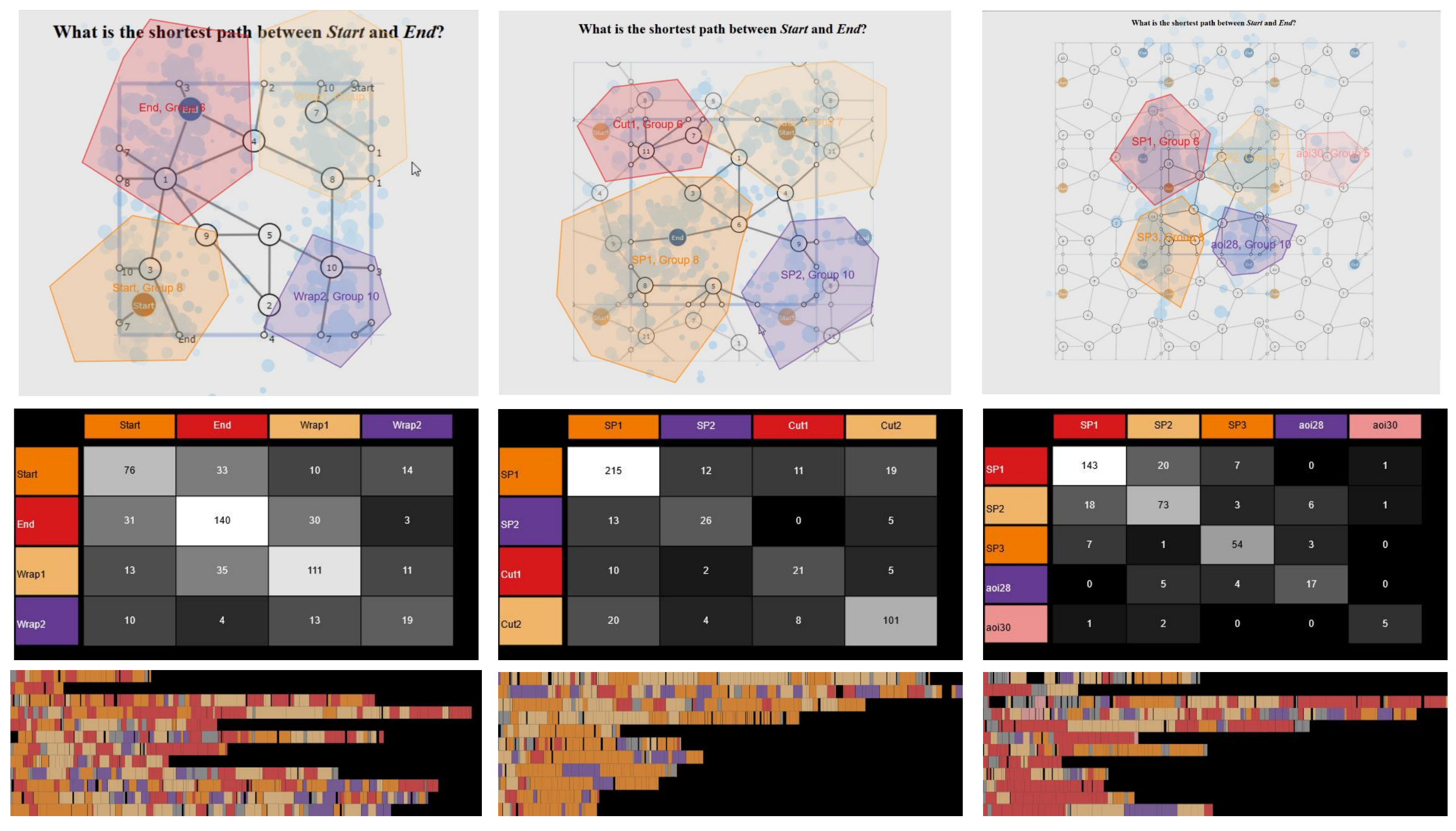}
    
    \Description[]{}
    \caption[]{AOI definitions and  transition matrices of a representative stimuli where the shortest path resides at corner wrapping for the shortest path task, across each participant. 
    \label{fig:aoi_transitions_corner_wrap}
    }
     \vspace{-1em}
\end{figure*}

\begin{figure}
    \centering
    \includegraphics[width=0.3\textwidth]{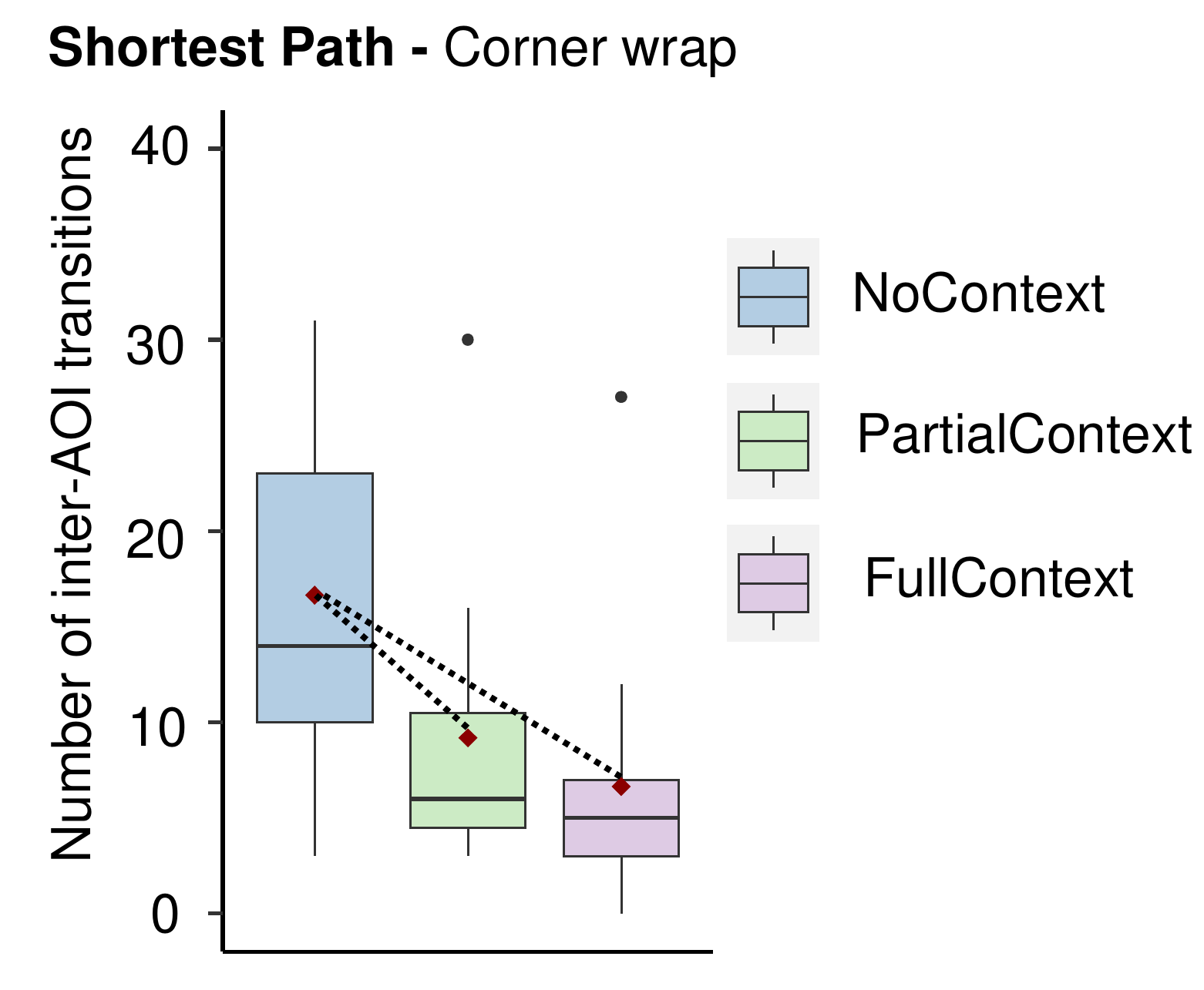}
    
    \Description[]{}
    \caption[]{Box plots of inter-AOI transitions of a representative stimuli (where the shortest path resides at corner wrapping) for the shortest path task, as shown in~\autoref{fig:aoi_transitions_corner_wrap}, across each participant. Significant results are indicated in dotted lines.  \label{fig:aoi_transitions_corner_wrap_stat}
    }
     \vspace{-1em}
\end{figure}
\subsection{Aggregated Fixation Distribution}
\label{sec:fixationdistribution}

\autoref{fig:teaser} shows the visualizations of aggregated fixation distribution for the \tshortestpath{} task for a representative stimulus, where the unique shortest path follows a top-bottom, left-right wrapping, or combination of them (i.e., corner wrapping). 
For \tnodelink{}, most fixations are concentrated on the center, forming one visual cluster of fixations over the shortest path between the start and the end node, as seen in~\autoref{fig:teaser}(a). 
Sole cluster of fixations is also found in all the other \tnodelink{} trials, independent of tasks (see the supplementary material). 
% Illustration of aggregated fixation distribution of other stimuli is shown in the supplementary material.

By contrast, \tnocontext{}, \tpartialcontext{}, and \tfullcontext{} show similarly clearer separation between clusters of fixations.
However, the relationship between these clusters vary across layout conditions. 
For \tnocontext{}, participants spent longer duration fixating at each of the start and the end node, along with their links to the boundaries where it discontinues. 
In addition, disjoint clusters of fixations at upper-right and bottom-right cover intermediate nodes of the shortest path, as seen in~\autoref{fig:teaser}(b), suggesting a relationship among them to connect the visual cuts wrapped across the boundaries.
This is different from \tpartialcontext{} and \tfullcontext{}. Two clusters of fixations cover the replication of unique shortest path were found for \tpartialcontext{} at the bottom-left and the bottom-right (\autoref{fig:teaser}(c)), and the same for \tfullcontext{} with three to four clusters (\autoref{fig:teaser}(d)). However, an additional dense fixations at upper-right in \tpartialcontext{} seems redundant since it is not possible to find the shortest path in adjacent tiles.
% To quantify the relationship between these clusters, we define AOIs and conduct comparative analysis in~\autoref{sec:comparative}.
% For \tpartialcontext{}, there is an apparent cluster separation where the bottom-left cluster covers a full copy of the unique shortest path, the upper-right cluster covers another, however, it apparently does not result in any full shortest path, implying an ineffective visual search here. As a result, a user needs to give up and start a new search (see scanpath analysis results below).
% We explore the relationship
% when inspecting the unique shortest path that wraps around. (2) 

% For \tpartialcontext{}, there is an apparent cluster separation where the bottom-left cluster covers a full copy of the unique shortest path, the upper-right cluster covers another, however, it apparently does not result in any full shortest path, implying an ineffective visual search here. As a result, a user needs to give up and start a new search (see scanpath analysis results below). 
For \tfullcontext{}, dense fixations were found surrounding the center and surrounding tiles, including a portion of corner tiles, but with decreasing fixations going outward dense maps, suggesting inefficient gaze allocation since participants are more in favor of center and adjacent tiles over the entire layout. 
% For \tfullcontext{}, fixations distributed outside the cluster of fixation points are mostly contributed by early fixations (shown in low saturated, i.e., light blue dots depicting participant's fixation sequence).  
% \qn{please change the order of (1), (2), (3), (4) in the order shown in~\autoref{fig:teaser}, or you have a reason for the current order? And current (1), (4) have "suggesting", "implying", but (2), (3) you only wrote what we see in the figure without drawing any implication? Why is that?}

%\tlinkcount{}
%results figure are available in supplementary materials.

% to gain an overview of the gaze behaviour.

% As the content of interests such as the unique shortest path, from the perspective of different wrapping directions or located at the centre without wrapping, we first visualize the overall attention distribution in density maps for each of top-bottom wrapping, left-right wrapping, and corner wrapping stimuli, by tasks, as seen in~\autoref{fig:teaser}.

% We employ a visual gaze analytics approach~\cite{kurzhals2017task} with a visual toolkit to explore spatiotemporal data~\footnote{URL is anonymized for peer review}. 

%

%
% We perform abovementioned multiway visual analysis through an exploratory visual eye tracking analytics toolkit, called webveta

\subsection{Reading Strategies Based on Scanpath Analysis}
\label{sec:scanpath}

% \begin{figure}
%     \centering
%      \subfigure[NoContext]{
%     \includegraphics[width=0.3\columnwidth]{figures/T14_NC_timeline.jpg}
%     }
%     \subfigure[PartialContext]{
%     \includegraphics[width=0.3\columnwidth]{figures/T14_PC_timelines(7).jpg}
%     }
%     \subfigure[FullContext]{
%     \includegraphics[width=0.3\columnwidth]{figures/T14_FC_timeline.jpg}
%     }
%     \subfigure[Number of AOI transitions]{
%     \includegraphics[width=0.3\columnwidth]{figures/aoi_transitions_corner_wrap_stat.pdf}
%     }
%     \Description[]{Box plot and spatial view of fixation of AOI percentage time where AOIs are defined by tiles (a-b) and by increasing distance from the centre cell (c-d), across medium and large stimuli of shortest path tasks}
%     \caption[]{Box plots of AOI transitions of a representative stimuli (where the shortest path resides at corner wrapping) for the shortest path task, as shown in~\autoref{fig:aoi_transitions_corner_wrap}, across each participant. Significant results are indicated in dotted lines. Confidence interval is 95\%. \label{fig:aoi_transitions_corner_wrap_stat}
%     }
%     \label{fig:aoi-percentage-time}
%     % \vspace{-1em}
% \end{figure}

%
% Fixation sequence is encoded by node colour in the overlay. 
% %
% Meanwhile, to gain a better understanding about a user's reading strategy, we also analyse individual participant' scanpaths of each layout in parallel with the density maps. 

\autoref{fig:scanpath_strategies} shows results of the analysis of scanpaths of the same stimulus as \autoref{sec:fixationdistribution}, ordered from left to right by their increasing task completion time.  
% The scanpath results of \tnodelink{} can be found in supplementary material. 
% The scanpaths represent users performing the graph exploration tasks correctly, ordered by increasing task completion time from the left to the right. 
 %   
% Visual search was found to often begin (yellow dots) and end (red dots) with either the start or the end node, regardless of layouts.
%
For \tnocontext{}, there is an increasing frequency of cross-checking the shortest path with low  (\autoref{fig:scanpath_strategies}(b)) to high frequency (\autoref{fig:scanpath_strategies}(d)). 
%
% For \tnocontext{}, scanpaths show a tendency to have more frequent transitions between AOIs than for \tpartialcontext{} and \tfullcontext{}. 
%
Furthermore, scanpaths were found switching between four corners with constant visual scan back to immediately starting regions, forming a frequent glance pattern between regions.  
We quantify the frequency of such visual search patterns and compare across each layout condition in~\autoref{sec:comparative}. 
For \tpartialcontext{}, \autoref{fig:scanpath_strategies}(g)--(i) show the scanpaths cover a full shortest path, however, there are also visual scan with a start or end node that does not lead to any shortest path due to the clipping of the repeated tiles, such as the orange start node near the upper-right corner, confirming the observation of redundant cluster of fixations in~\autoref{sec:fixationdistribution}.  
% why not then explicitly says "Thanks to this pattern from the explorative study, we propose to quantify it for further...., right now it feels like you are writing the methodology section
For \tfullcontext{}, \autoref{fig:scanpath_strategies}(l)--(n) show one to two effective searches out of all $3\times3$ tiles.
% to quantify..-> sounds like methodology.....to do something, not as a results of the explorative..

% Overall, \tnocontext{}-\tshortestpath{} has significantly more glance saccades than \tnodelink{}. We leave its statistical testing results to~\autoref{sec:comparative}.

% We also note that users' reading \tnocontext{} has more than one cluster of fixations, as shown in second row top-bottom, left-right, corner wrapped stimuli, while in no-wrap, \tnocontext{} has just one cluster similar to traditional \tnodelink{} layout. 

% \tnocontext{} layout adopts strategy of more frequent long range glancing after immediately fixated areas, and has significantly more transitions between two or more density kernels (clusters of fixation points) than all the other layout conditions.
% We suggest to automatically centre content of interest to avoid the visual cut.
% In partial context, frequent visual searches over redundant areas are identified. 
% In full-context the visual search concentrates on one or two complete shortest paths leading to more effective performance in time, but the large screen space is quite inefficient based on scanpaths. 

Our findings confirm existing path-following reading strategies in the study by \citet{netzel2017user}, there are participants showing scan once, multiple rounds, inspecting other possible paths between origin and destination. 
%
% Scanpaths of other stimuli of various wrapping directions, including top-bottom, and left-right wrapping show similar patterns to the results presented in this subsection; see the supplementary material. 

% \qn{Personally, I find it's hard to grab the order of findings (1), (2), .., (5), and link among them. Sometimes (1) For Nocontext, but then (2) also "for nocontext" as starting, then (3) doesn't mention any type of stimuli, (4) go back to "for partialcontext", then (5) for "fullcontext". I'm confused. How about merging (1), (2) (Nocontext fact compared with others)? How about (3)? About fig-crossing, (1) you linked to Fig 3a - 3d, but not (5), for instance, which figs I'm supposed to look at for (5), the whole third row?}

% we choose one representitative stimuli where horizontal, vertical, and corner wrapping are involved in the same graph, laid out in three representations. 
% From the users' scanpaths, we also find distinguishable reading strategies. 

\subsection{Comparative Statistical Analysis Based on Gaze Metrics}
\label{sec:comparative}
Inspired by aggregated fixation distribution (\autoref{sec:fixationdistribution}) and scanpath strategies results (\autoref{sec:scanpath}), we  quantify the patterns with metrics and conduct statistical analysis based on metrics.

To quantify the glance patterns, we classify saccades into three categories: short (less than 100 pixels), long, and glance saccades. 
% hey hey hey, then what happens after the quantification? It feels like you are mixing between methodology with results in this subsection.
% I would say this is told in the beginning. We perform exploratory analysis, and then inspired from that, we establish what we quantify for comparative analysis. Is that message clear when you read through starting from Sect.4?
A glance saccade is defined as saccades from a fixation at time index $t-1$ to one at time $t$ and fall back to one at time $t+1$ so that the Euclidean distance between fixations at $t-1$ and $t+1$ is less than one third of the distance between fixations at time $t$ and $t+1$, i.e., a glance saccade occurs when gaze moves back and forth. 
A similar approach has also been used ~\cite{goodwin2022veta}. 

% We perform the analysis by aggregating medium and large graphs of a representative stimulus.
We use the \emph{Friedman} test to evaluate significant differences between layout conditions where the dependent variables (shown below) did not meet the normality assumption. We use the \emph{Nemenyi} post-hoc test for pairwise comparisons.
Confidence intervals indicate 95\% confidence for mean values.
% Significance values are reported for $p<.05(∗)$,
% $p<.01(∗∗)$, and $p<.001(∗∗∗)$, abbreviated by the number of stars
% in parenthesis.
% Numbers in parentheses indicate mean values and 95\% confidence intervals (CI).

Results of saccade frequencies of \tshortestpath{} for medium and large stimuli across layouts are shown in~\autoref{fig:number-of-saccades}. 
Overall, there is an effect of the number of short ($\chi^2(3)=30.47, p<.001$), long ($\chi^2(3)=9.14, p=.027$), and glance saccades ($\chi^2(3)=10.77, p=.013$) on layout conditions. Post-hoc analysis shows that \tnodelink{} has significantly less short saccades than \tnocontext{} ($p<.001$), \tpartialcontext{} ($p<.001$), and \tfullcontext{} ($p=.0014$). 
\tnodelink{} has significantly less long saccades than \tnocontext{} ($p=.034$), and a weak trend to have less long saccades than \tpartialcontext{} ($p=.073$).
\tnocontext{} layout has significantly more  glance saccades than \tpartialcontext{} ($p=.041$), and a weak trend to have more glance saccades than \tfullcontext{} ($p=.061$). 

Furthermore, similar to ~\citet{EyeTrackingTree2011}, we define AOIs to investigate the relationships between clusters of fixations between the visual clusters of fixations identified in density maps for each layout, as seen in \autoref{fig:aoi_transitions_corner_wrap}. 
% These AOIs are visually similar to automatically created AOIs from Tobii Pro Lab with a distance clustering threshold of 12\%. 
% These AOIs are similar in shape and size to the AOIs automatically created by Tobii Pro Lab based on a distance clustering metric (see the supplementary material).
% Furthermore, we find for \tnocontext{}, there are more glance saccades than \tpartialcontext{} and \tfullcontext{}. 
% We manipulate AOI definition for each layout while visually inspecting the hit-any-AOI-rate (HAAR) metric using the visual analysis toolkit, until the HAAR metric for each participant is above 80\%, and there is no significant difference in HAAR metric between layout conditions (see the box plots in supplementary material).
% This metric characterizes the impact of AOI uncertainty on stimuli~\cite{wang2022impact}.
% Note that for manually created AOIs, we did a significant statistical testing of its definitions on HAAR across layout condition
% we export this metric for every participant per condition for a Friedman statistics testing. 
% The result shows there is no significant difference in HAAR across conditions (see supplementary material).
The comparison of the number of inter-AOI transitions indicates that overall there is an effect of inter-AOI transitions on layout conditions ($\chi^2(2)=8.97, p=.011$). Posthoc analysis shows that \tnocontext{} has significantly more inter-AOI transitions than \tpartialcontext{} ($p=.021$) and \tfullcontext{} ($p=.038$), as seen in~\autoref{fig:aoi_transitions_corner_wrap_stat}. 
A number of participants were found sharing similar investigatory patterns. They are grouped in the scarfplot based on their pairwise visual similarity AOI transitions~\cite{kurzhals2014iseecube} (\autoref{fig:aoi_transitions_corner_wrap_stat}(bottom)).

Both results indicate that (1) the low-performance of \tnocontext{} could be a result of increasing mental effort to perform more frequent saccades and inter-AOI transitions when the content of interest wraps around. In particular, long-range glance saccades are used more often than the other layouts to connect visual cuts of links and paths across the boundaries. 
(2) While there are similar number of clusters of fixations for each torus-based layout, there is different strategic use of these clusters. There are  more inter-cluster transitions in \tnocontext{} than in \tpartialcontext{} and \tfullcontext{}, while the latter has more self-transitions. 
% Fig 6, the partial and the full don't have significant difference, but there is tendency that full is better than partial? No, yes? Because the writing right now is like nocontext vs. the rest.. :) 

% Scanpaths of other stimuli of various wrapping directions, including top-bottom, and left-right wrapping show similar patterns to the results presented in this subsection; see the supplementary material.

%% file: sections/conclusion.tex
\section{Discussion and Conclusion}
%We address our research questions in two ways. First, 
We studied different reading strategies in wrapped node-link layout representations using eye tracking.
Our results indicate that there are distinguishable reading strategies by aggregated fixation distributions, scanpaths, and comparative gaze metrics.  
% Hi man, can we remove 3.3?
In addition, we provide further evidence that the low-performance of \tpartialcontext{} from the prior study is caused by ineffective visual searches where the path continuation could not be followed without starting a new search in repeated tiles. Furthermore, low-performance of \tnocontext{} is caused by more frequent glance saccades, as well as  more transitions between two or more clusters of fixation points. 
This complements existing findings for task performance by \citet{chen2020doughnets}.
% These results address research questions qualitatively and quantitatively. 

Our results provide implications not only for future torus-based study design for node-link diagrams, but also for the utilization of different levels of torus-based layouts. 
These implications might provide a starting point for future studies of other types of wrappable visualizations~\cite{chen2022s}, such as cyclic time series where inspecting trends across the visual cuts has been identified as a major problem in~\cite{chen2021rotate}. 
%
%\qn{long sentence, can be reduced} hey hey hey please go to the study.tex, ok, please finish here first then
%
% For instance, for both \tnocontext{} and \tpartialcontext{}, we recommend the contents should be centered to avoid the need to either frequently scan the boundaries to connect the visual cut (\tnocontext{}). 
A new layout could be derived from a balance of \tpartialcontext{} and \tfullcontext{}, where a more space-efficient wrapped visualization is suggested to clip the outer-most repetition rather than at the four corner tiles, to cover effective visual search (informed by the gaze allocation results in this paper as well as the supplementary material).

\rev{Similar to existing eye tracking studies on node-link visualizations \cite{huang2009graph,burch2013visual}, our analysis approach has similar limitations. We based our visual search behavior analysis on the same type of stimuli aggregated across all the participants. 
Accordingly, we focused on reporting results from  representative stimuli of each layout condition, where their wrapping conditions of the shortest path is on the same basis of a corner wrapping. 
% Note that our results should be taken carefully for this wrapping variable only. 
While we infer a general reading strategies from this particular corner wrapped condition, it is a combination of top--bottom and left--right wrap. 
Future work can extend it to considering a full range of wrapping conditions.
We include aggregated density map visualizations of other wrapping conditions in the supplementary material, including top--bottom wrap, left--right wrap, and no--wrap.}

\rev{Another} limitation of this work is the small sample size (13 participants). 
Future studies could be conducted with more subjects to fill this gap. 
Another direction for future research is to investigate whether the visual clutter reduction affordable in a torus-based network layout and the design recommendations informed by gaze allocations might lead to more usable and effective graph layouts compared with traditional non-wrapped representations.

%% file: main.bbl

\begin{thebibliography}{23}

%%% the Web address.

\ifx \showCODEN    \undefined \def \showCODEN     #1{\unskip}     \fi
\ifx \showDOI      \undefined \def \showDOI       #1{#1}\fi
\ifx \showISBNx    \undefined \def \showISBNx     #1{\unskip}     \fi
\ifx \showISBNxiii \undefined \def \showISBNxiii  #1{\unskip}     \fi
\ifx \showISSN     \undefined \def \showISSN      #1{\unskip}     \fi
\ifx \showLCCN     \undefined \def \showLCCN      #1{\unskip}     \fi
\ifx \shownote     \undefined \def \shownote      #1{#1}          \fi
\ifx \showarticletitle \undefined \def \showarticletitle #1{#1}   \fi
\ifx \showURL      \undefined \def \showURL       {\relax}        \fi
% invisible to TeX
\providecommand\bibfield[2]{#2}
\providecommand\bibinfo[2]{#2}
\providecommand\natexlab[1]{#1}
\providecommand\showeprint[2][]{arXiv:#2}

\bibitem[Andrienko et~al\mbox{.}(2012)]%
        {Andrienko:2012:VAM}
\bibfield{author}{\bibinfo{person}{Gennady~L. Andrienko},
  \bibinfo{person}{Natalia~V. Andrienko}, \bibinfo{person}{Michael Burch},
  {and} \bibinfo{person}{Daniel Weiskopf}.} \bibinfo{year}{2012}\natexlab{}.
\newblock \showarticletitle{Visual Analytics Methodology for Eye Movement
  Studies}.
\newblock \bibinfo{journal}{\emph{{IEEE} Transactions on Visualization and
  Computer Graphics}} \bibinfo{volume}{18}, \bibinfo{number}{12}
  (\bibinfo{year}{2012}), \bibinfo{pages}{2889--2898}.
\newblock
\urldef\tempurl%
\url{https://doi.org/10.1109/TVCG.2012.276}
\showDOI{\tempurl}


\bibitem[Blascheck et~al\mbox{.}(2017)]%
        {Tanja2017NodeLink}
\bibfield{author}{\bibinfo{person}{Tanja Blascheck}, \bibinfo{person}{Markus
  Schweizer}, \bibinfo{person}{Fabian Beck}, {and} \bibinfo{person}{Thomas
  Ertl}.} \bibinfo{year}{2017}\natexlab{}.
\newblock \showarticletitle{Visual Comparison of Eye Movement Patterns}.
\newblock \bibinfo{journal}{\emph{Computer Graphics Forum}}
  \bibinfo{volume}{36}, \bibinfo{number}{3} (\bibinfo{year}{2017}),
  \bibinfo{pages}{87--97}.
\newblock
\urldef\tempurl%
\url{https://doi.org/10.1111/cgf.13170}
\showDOI{\tempurl}


\bibitem[Burch et~al\mbox{.}(2013)]%
        {burch2013visual}
\bibfield{author}{\bibinfo{person}{Michael Burch}, \bibinfo{person}{Gennady
  Andrienko}, \bibinfo{person}{Natalia Andrienko}, \bibinfo{person}{Markus
  H{\"o}ferlin}, \bibinfo{person}{Michael Raschke}, {and}
  \bibinfo{person}{Daniel Weiskopf}.} \bibinfo{year}{2013}\natexlab{}.
\newblock \showarticletitle{Visual Task Solution Strategies in Tree Diagrams}.
  In \bibinfo{booktitle}{\emph{2013 IEEE Pacific Visualization Symposium
  (PacificVis)}}. \bibinfo{pages}{169--176}.
\newblock
\urldef\tempurl%
\url{https://doi.org/10.1109/PacificVis.2013.6596142}
\showDOI{\tempurl}


\bibitem[Burch et~al\mbox{.}(2011)]%
        {EyeTrackingTree2011}
\bibfield{author}{\bibinfo{person}{Michael Burch}, \bibinfo{person}{Natalia
  Konevtsova}, \bibinfo{person}{Julian Heinrich}, \bibinfo{person}{Markus
  Höferlin}, {and} \bibinfo{person}{Daniel Weiskopf}.}
  \bibinfo{year}{2011}\natexlab{}.
\newblock \showarticletitle{Evaluation of Traditional, Orthogonal, and Radial
  Tree Diagrams by an Eye Tracking Study}.
\newblock \bibinfo{journal}{\emph{IEEE Transactions on Visualization and
  Computer Graphics}} \bibinfo{volume}{17}, \bibinfo{number}{12}
  (\bibinfo{year}{2011}), \bibinfo{pages}{2440--2448}.
\newblock
\urldef\tempurl%
\url{https://doi.org/10.1109/TVCG.2011.193}
\showDOI{\tempurl}


\bibitem[Chang et~al\mbox{.}(2017)]%
        {chang2017evaluating}
\bibfield{author}{\bibinfo{person}{Chunlei Chang}, \bibinfo{person}{Benjamin
  Bach}, \bibinfo{person}{Tim Dwyer}, {and} \bibinfo{person}{Kim Marriott}.}
  \bibinfo{year}{2017}\natexlab{}.
\newblock \showarticletitle{Evaluating Perceptually Complementary Views for
  Network Exploration Tasks}. In \bibinfo{booktitle}{\emph{Proceedings of the
  2017 CHI Conference on Human Factors in Computing Systems}}.
  \bibinfo{pages}{1397--1407}.
\newblock
\urldef\tempurl%
\url{https://doi.org/10.1145/3025453.3026024}
\showDOI{\tempurl}


\bibitem[Chen(2022)]%
        {chen2022s}
\bibfield{author}{\bibinfo{person}{Kun-Ting Chen}.}
  \bibinfo{year}{2022}\natexlab{}.
\newblock \showarticletitle{It's a Wrap! Visualisations that Wrap Around
  Cylindrical, Toroidal, or Spherical Topologies}.
\newblock \bibinfo{journal}{\emph{arXiv preprint arXiv:2209.13251}}
  (\bibinfo{year}{2022}).
\newblock


\bibitem[Chen et~al\mbox{.}(2021a)]%
        {chen2021sa}
\bibfield{author}{\bibinfo{person}{Kun-Ting Chen}, \bibinfo{person}{Tim Dwyer},
  \bibinfo{person}{Benjamin Bach}, {and} \bibinfo{person}{Kim Marriott}.}
  \bibinfo{year}{2021}\natexlab{a}.
\newblock \showarticletitle{It’s a Wrap: Toroidal Wrapping of Network
  Visualisations Supports Cluster Understanding Tasks}. In
  \bibinfo{booktitle}{\emph{Proceedings of the 2021 CHI Conference on Human
  Factors in Computing Systems}}. Article \bibinfo{articleno}{465},
  \bibinfo{numpages}{12}~pages.
\newblock
\urldef\tempurl%
\url{https://doi.org/10.1145/3411764.3445439}
\showDOI{\tempurl}


\bibitem[Chen et~al\mbox{.}(2021b)]%
        {chen2021rotate}
\bibfield{author}{\bibinfo{person}{Kun-Ting Chen}, \bibinfo{person}{Tim Dwyer},
  \bibinfo{person}{Benjamin Bach}, {and} \bibinfo{person}{Kim Marriott}.}
  \bibinfo{year}{2021}\natexlab{b}.
\newblock \showarticletitle{Rotate or Wrap? Interactive Visualisations of
  Cyclical Data on Cylindrical or Toroidal Topologies}.
\newblock \bibinfo{journal}{\emph{IEEE Transactions on Visualization and
  Computer Graphics}} \bibinfo{volume}{28}, \bibinfo{number}{1}
  (\bibinfo{year}{2021}), \bibinfo{pages}{727--736}.
\newblock
\urldef\tempurl%
\url{https://doi.org/10.1109/TVCG.2021.3114693}
\showDOI{\tempurl}


\bibitem[Chen et~al\mbox{.}(2020)]%
        {chen2020doughnets}
\bibfield{author}{\bibinfo{person}{Kun-Ting Chen}, \bibinfo{person}{Tim Dwyer},
  \bibinfo{person}{Kim Marriott}, {and} \bibinfo{person}{Benjamin Bach}.}
  \bibinfo{year}{2020}\natexlab{}.
\newblock \showarticletitle{DoughNets: Visualising Networks Using Torus
  Wrapping}. In \bibinfo{booktitle}{\emph{Proceedings of the 2020 CHI
  Conference on Human Factors in Computing Systems}}. Article
  \bibinfo{articleno}{53}, \bibinfo{numpages}{11}~pages.
\newblock
\urldef\tempurl%
\url{https://doi.org/10.1145/3313831.3376180}
\showDOI{\tempurl}


\bibitem[Chen et~al\mbox{.}(2022)]%
        {chen2022gan}
\bibfield{author}{\bibinfo{person}{Kun-Ting Chen}, \bibinfo{person}{Tim Dwyer},
  \bibinfo{person}{Yalong Yang}, \bibinfo{person}{Benjamin Bach}, {and}
  \bibinfo{person}{Kim Marriott}.} \bibinfo{year}{2022}\natexlab{}.
\newblock \showarticletitle{GAN'SDA Wrap: Geographic And Network Structured
  DAta on surfaces that Wrap around}. In \bibinfo{booktitle}{\emph{Proceedings
  of the 2022 CHI Conference on Human Factors in Computing Systems}}. Article
  \bibinfo{articleno}{135}, \bibinfo{numpages}{16}~pages.
\newblock
\urldef\tempurl%
\url{https://doi.org/10.1145/3491102.3501928}
\showDOI{\tempurl}


\bibitem[Chen et~al\mbox{.}(2023)]%
        {chen2023gazealytics}
\bibfield{author}{\bibinfo{person}{Kun-Ting Chen}, \bibinfo{person}{Arnaud
  Prouzeau}, \bibinfo{person}{Joshua Langmead}, \bibinfo{person}{Ryan~T
  Whitelock-Jones}, \bibinfo{person}{Lee Lawrence}, \bibinfo{person}{Tim
  Dwyer}, \bibinfo{person}{Christophe Hurter}, \bibinfo{person}{Daniel
  Weiskopf}, {and} \bibinfo{person}{Sarah Goodwin}.}
  \bibinfo{year}{2023}\natexlab{}.
\newblock \bibinfo{title}{Gazealytics: A Unified and Flexible Visual Toolkit
  for Exploratory and Comparative Gaze Analysis}.
\newblock
\newblock
\showeprint[arxiv]{2303.17202}~[cs.HC]


\bibitem[Goodwin et~al\mbox{.}(2022)]%
        {goodwin2022veta}
\bibfield{author}{\bibinfo{person}{Sarah Goodwin}, \bibinfo{person}{Arnaud
  Prouzeau}, \bibinfo{person}{Ryan Whitelock-Jones},
  \bibinfo{person}{Christophe Hurter}, \bibinfo{person}{Lee Lawrence},
  \bibinfo{person}{Umair Afzal}, {and} \bibinfo{person}{Tim Dwyer}.}
  \bibinfo{year}{2022}\natexlab{}.
\newblock \showarticletitle{VETA: Visual eye-tracking analytics for the
  exploration of gaze patterns and behaviours}.
\newblock \bibinfo{journal}{\emph{Visual Informatics}} \bibinfo{volume}{6},
  \bibinfo{number}{2} (\bibinfo{year}{2022}), \bibinfo{pages}{1--13}.
\newblock


\bibitem[Huang(2007)]%
        {GraphLayoutEffectEyeTracking2007}
\bibfield{author}{\bibinfo{person}{Weidong Huang}.}
  \bibinfo{year}{2007}\natexlab{}.
\newblock \showarticletitle{Using Eye Tracking to Investigate Graph Layout
  Effects}. In \bibinfo{booktitle}{\emph{Proceedings of the Asia-Pacific
  Symposium on Visualisation 2007}}. \bibinfo{pages}{97--100}.
\newblock
\urldef\tempurl%
\url{https://doi.org/10.1109/APVIS.2007.329282}
\showDOI{\tempurl}


\bibitem[Huang(2013)]%
        {Huang2013}
\bibfield{author}{\bibinfo{person}{Weidong Huang}.}
  \bibinfo{year}{2013}\natexlab{}.
\newblock \showarticletitle{Establishing aesthetics based on human graph
  reading behavior: two eye tracking studies}.
\newblock \bibinfo{journal}{\emph{Personal and Ubiquitous Computing}}
  \bibinfo{volume}{17}, \bibinfo{number}{1} (\bibinfo{year}{2013}),
  \bibinfo{pages}{93--105}.
\newblock
\showISSN{1617-4917}
\urldef\tempurl%
\url{https://doi.org/10.1007/s00779-011-0473-2}
\showDOI{\tempurl}


\bibitem[Huang et~al\mbox{.}(2009)]%
        {huang2009graph}
\bibfield{author}{\bibinfo{person}{Weidong Huang}, \bibinfo{person}{Peter
  Eades}, {and} \bibinfo{person}{Seok-Hee Hong}.}
  \bibinfo{year}{2009}\natexlab{}.
\newblock \showarticletitle{A Graph Reading Behavior: Geodesic-Path Tendency}.
  In \bibinfo{booktitle}{\emph{Proceedings of the IEEE Pacific Visualization
  Symposium}}. \bibinfo{pages}{137--144}.
\newblock
\urldef\tempurl%
\url{https://doi.org/10.1109/PACIFICVIS.2009.4906848}
\showDOI{\tempurl}


\bibitem[Jianu et~al\mbox{.}(2014)]%
        {DisplayGroupInfoNodeLink2014}
\bibfield{author}{\bibinfo{person}{Radu Jianu}, \bibinfo{person}{Adrian Rusu},
  \bibinfo{person}{Yifan Hu}, {and} \bibinfo{person}{Douglas Taggart}.}
  \bibinfo{year}{2014}\natexlab{}.
\newblock \showarticletitle{How to Display Group Information on Node-Link
  Diagrams: An Evaluation}.
\newblock \bibinfo{journal}{\emph{IEEE Transactions on Visualization and
  Computer Graphics}} \bibinfo{volume}{20}, \bibinfo{number}{11}
  (\bibinfo{year}{2014}), \bibinfo{pages}{1530--1541}.
\newblock
\urldef\tempurl%
\url{https://doi.org/10.1109/TVCG.2014.2315995}
\showDOI{\tempurl}


\bibitem[Kumar et~al\mbox{.}(2019)]%
        {kumar2019clustered}
\bibfield{author}{\bibinfo{person}{Ayush Kumar}, \bibinfo{person}{Neil
  Timmermans}, \bibinfo{person}{Michael Burch}, {and} \bibinfo{person}{Klaus
  Mueller}.} \bibinfo{year}{2019}\natexlab{}.
\newblock \showarticletitle{Clustered Eye Movement Similarity Matrices}. In
  \bibinfo{booktitle}{\emph{Proceedings of the 11th ACM Symposium on Eye
  Tracking Research \& Applications}}. Article \bibinfo{articleno}{82},
  \bibinfo{numpages}{9}~pages.
\newblock
\urldef\tempurl%
\url{https://doi.org/10.1145/3317958.3319811}
\showDOI{\tempurl}


\bibitem[Kurzhals et~al\mbox{.}(2017)]%
        {kurzhals2017task}
\bibfield{author}{\bibinfo{person}{Kuno Kurzhals}, \bibinfo{person}{Michael
  Burch}, \bibinfo{person}{Tanja Blascheck}, \bibinfo{person}{Gennady
  Andrienko}, \bibinfo{person}{Natalia Andrienko}, {and}
  \bibinfo{person}{Daniel Weiskopf}.} \bibinfo{year}{2017}\natexlab{}.
\newblock \showarticletitle{A Task-Based View on the Visual Analysis of
  Eye-Tracking Data}.
\newblock In \bibinfo{booktitle}{\emph{Eye Tracking and Visualization}},
  \bibfield{editor}{\bibinfo{person}{Michael Burch}, \bibinfo{person}{Lewis
  Chuang}, \bibinfo{person}{Brian Fisher}, \bibinfo{person}{Albrecht Schmidt},
  {and} \bibinfo{person}{Daniel Weiskopf}} (Eds.). \bibinfo{publisher}{Springer
  International Publishing}, \bibinfo{address}{Cham}, \bibinfo{pages}{3--22}.
\newblock
\showISBNx{978-3-319-47024-5}


\bibitem[Kurzhals et~al\mbox{.}(2014)]%
        {kurzhals2014iseecube}
\bibfield{author}{\bibinfo{person}{Kuno Kurzhals}, \bibinfo{person}{Florian
  Heimerl}, {and} \bibinfo{person}{Daniel Weiskopf}.}
  \bibinfo{year}{2014}\natexlab{}.
\newblock \showarticletitle{ISeeCube: Visual analysis of gaze data for video}.
  In \bibinfo{booktitle}{\emph{Proceedings of the Symposium on Eye Tracking
  Research and Applications}}. \bibinfo{pages}{43--50}.
\newblock


\bibitem[Munzner(2014)]%
        {munzner2014visualization}
\bibfield{author}{\bibinfo{person}{Tamara Munzner}.}
  \bibinfo{year}{2014}\natexlab{}.
\newblock \bibinfo{booktitle}{\emph{Visualization Analysis and Design}}.
\newblock \bibinfo{publisher}{CRC Press}, \bibinfo{address}{Boca Raton, FL}.
\newblock


\bibitem[Netzel et~al\mbox{.}(2017)]%
        {netzel2017user}
\bibfield{author}{\bibinfo{person}{Rudolf Netzel}, \bibinfo{person}{Bettina
  Ohlhausen}, \bibinfo{person}{Kuno Kurzhals}, \bibinfo{person}{Robin Woods},
  \bibinfo{person}{Michael Burch}, {and} \bibinfo{person}{Daniel Weiskopf}.}
  \bibinfo{year}{2017}\natexlab{}.
\newblock \showarticletitle{User erformance and reading strategies for metro
  maps: An eye tracking study}.
\newblock \bibinfo{journal}{\emph{Spatial Cognition \& Computation}}
  \bibinfo{volume}{17}, \bibinfo{number}{1-2} (\bibinfo{year}{2017}),
  \bibinfo{pages}{39--64}.
\newblock
\urldef\tempurl%
\url{https://doi.org/10.1080/13875868.2016.1226839}
\showDOI{\tempurl}


\bibitem[Pohl et~al\mbox{.}(2009)]%
        {pohl2009comparing}
\bibfield{author}{\bibinfo{person}{Mathias Pohl}, \bibinfo{person}{Markus
  Schmitt}, {and} \bibinfo{person}{Stephan Diehl}.}
  \bibinfo{year}{2009}\natexlab{}.
\newblock \showarticletitle{Comparing the Readability of Graph Layouts using
  Eyetracking and Task-oriented Analysis}. In
  \bibinfo{booktitle}{\emph{Proceedings of Computational Aesthetics in
  Graphics, Visualization, and Imaging}}. \bibinfo{pages}{49--56}.
\newblock
\urldef\tempurl%
\url{https://doi.org/10.2312/COMPAESTH/COMPAESTH09/049-056}
\showDOI{\tempurl}


\bibitem[Salvucci and Goldberg(2000)]%
        {salvucci2000identifying}
\bibfield{author}{\bibinfo{person}{Dario~D. Salvucci} {and}
  \bibinfo{person}{Joseph~H. Goldberg}.} \bibinfo{year}{2000}\natexlab{}.
\newblock \showarticletitle{Identifying Fixations and Saccades in Eye-Tracking
  Protocols}. In \bibinfo{booktitle}{\emph{Proceedings of the 2000 Symposium on
  Eye Tracking Research \& Applications}}. \bibinfo{pages}{71--78}.
\newblock
\urldef\tempurl%
\url{https://doi.org/10.1145/355017.355028}
\showDOI{\tempurl}


\end{thebibliography}
